\documentclass[onecolumn,showpacs,preprintnumbers,amsmath,amssymb,times,pre]{revtex4}

\usepackage{graphicx}
\usepackage{dcolumn}
\usepackage{bm}
\usepackage{amsmath}
\usepackage{color}

\begin{document}

\title{Promoting cooperation by punishing minority}

\author{Han-Xin Yang$^{1}$}\email{yanghanxin001@163.com}
\author{Xiaojie Chen$^{2}$}

\affiliation{$^{1}$Department of Physics, Fuzhou University, Fuzhou
350116, China\\$^{2}$School of Mathematical Sciences, University of
Electronic Science and Technology of China, Chengdu 611731, China }

\begin{abstract}
Punishment is an effective way to sustain cooperation among selfish
individuals. In most of previous studies, objects of punishment are
set to be defectors. In this paper, we propose a mechanism of
punishment, in which individuals with the majority strategy will
punish those with the minority strategy in a public goods game
group. Both theoretical analysis and simulation show that the
cooperation level can be greatly enhanced by punishing minority. For
no punishment or small values of punishment fine, the fraction of
cooperators continuously increases with the multiplication factor.
However, for large values of punishment fine, there exists a
critical value of multiplication factor, at which the fraction of
cooperators suddenly jumps from 0 to 1. The density of different
types of groups is also studied.
\end{abstract}

\maketitle



\section{Introduction}

To understand the emergence of cooperative behavior among selfish
individuals, researchers have considered various
mechanisms~\cite{rev1,rev2,rev3}, such as network
reciprocity~\cite{1,2,2.1}, voluntary participation~\cite{3,4},
teaching activity~\cite{5,7,8}, social diversity~\cite{9},
migration~\cite{11,13}, chaotic payoff variations~\cite{14},
extortion~\cite{15,16,17}, reputation~\cite{18},
memory~\cite{19,20}, diverse activity patterns~\cite{21}, the
coevolution setup~\cite{22}, onymity~\cite{23} and so on.

So far, punishment has been proved to be an effective way to enforce
cooperative behavior and various mechanisms of punishment have been
proposed~\cite{the1,the3,the4,the5,the6,the7,the8}. Szolnoki $et$
$al$ found that the impact of pool punishment on the evolution of
cooperation in structured populations is significantly different
from that reported previously for peer punishment~\cite{Szolnoki}.
Perc and Szolnoki proposed an adaptive punishment that can promote
public cooperation through the invigoration of spatial reciprocity,
the prevention of the emergence of cyclic dominance, or the
provision of competitive advantages to those that sanction
antisocial behavior~\cite{Perc}. Chen $et$ $al$ showed that sharing
the responsibility to sanction defectors rather than relying on
certain individuals to do so permanently can solve the problem of
costly punishment~\cite{Chen}. Cui and Wu demonstrated that the
presence of selfish punishment with avoiding mechanism can help
individuals out of both first-order and second-order social
dilemma~\cite{Wu}.

In previous studies, objects of punishment are individuals who hold
a specific strategy (usually is deemed to be defection). However,
the punished strategy may not be fixed but depends on the
surrounding environment, e.g., on neighbors' strategies.
Psychological experiments have demonstrated that, an individual
tends to follow the majority in behavior or opinion. There exist
some psychological or financial punishments of being
minority~\cite{dissent1,dissent2}. Based on the above consideration,
we propose a mechanism of punishment in which individuals with the
majority strategy will punish those with the minority strategy in a
group. Utilizing the public goods game (PGG) as a prototypical model
of group interaction, we find that cooperation can be greatly
promoted by punishing minority.

\section{Model}\label{sec:model}

Our model is described as follows.

Individuals are located on a $1000\times 1000$ square lattice with
periodic boundary conditions. Every individual occupies a lattice
point and has four neighboring points. A PGG group is composed of a
sponsor and its four neighbors. Thus the size of each PGG group is
5. Each individual $i$ participates in five different PGG groups
organized by $i$ and its four neighbors respectively.

At each time step, every cooperator contributes a unit cost to each
involved PGG group. Defectors invest nothing. The total cost of a
group is multiplied by a factor $r$, and is then redistributed
uniformly to all the five players in this group. In every PGG group,
individuals with the majority strategy will punish those with the
minority strategy. Each minority in the group is punished with a
fine $\alpha$. Following the previous study~\cite{Chen}, we assume
that punishers equally share the associated costs. If cooperators
are majority and defectors are minority in the group, then each
cooperator pays a cost $\alpha(5-n_{c})/n_{c}$, where $n_{c}$ is the
number of cooperators in the group. Oppositely, if defectors are
majority and cooperators are minority in the group, then each
defector pays a cost $\alpha n_{c}/(5-n_{c})$.

We denote $i$'s strategy as $s_{i}=1$ for cooperation and $s_{i}=0$
for defection. The payoff that player $i$ gains from the group
organized by player $j$ is calculated by the following equations:
\begin{equation}
if \ n_{c}=0 \ or \ 5, \ \Pi_{i}^{j}=-s_{i}+\frac{rn_{c}}{5};
\label{eq1}
\end{equation}
\begin{equation}
if \ 2<n_{c}<5 \ and \ s_{i}=1, \
\Pi_{i}^{j}=-1+\frac{rn_{c}}{5}-\frac{\alpha(5-n_{c})}{n_{c}};
\label{eq2}
\end{equation}
\begin{equation}
if \ 2<n_{c}<5 \ and \ s_{i}=0, \
\Pi_{i}^{j}=\frac{rn_{c}}{5}-\alpha; \label{eq3}
\end{equation}
\begin{equation}
if \ 0<n_{c}<3 \ and \ s_{i}=1, \
\Pi_{i}^{j}=-1+\frac{rn_{c}}{5}-\alpha; \label{eq4}
\end{equation}
\begin{equation}
if \ 0<n_{c}<3 \ and \ s_{i}=0, \
\Pi_{i}^{j}=\frac{rn_{c}}{5}-\frac{\alpha n_{c}}{5-n_{c}}.
\label{eq5}
\end{equation}
Equation~(\ref{eq1}) means that there is no punishment when the
group is occupied by full cooperators or full defectors. For
equations~(\ref{eq2}) and~(\ref{eq3}), cooperators punish defectors
in the group. While for equations~(\ref{eq4}) and~(\ref{eq5}),
defectors punish cooperators in the group. The total payoff of the
player $i$ is calculated by
\begin{equation}
P_{i}=\sum_{j\epsilon\Omega_{i}} \Pi_{i}^{j},\label{eq6}
\end{equation}
where $\Omega_{i}$ denotes the community of neighbors of $i$ and
itself.

Initially, cooperators and defectors are randomly distributed with
the equal probability 0.5. Individuals asynchronously update their
strategies in a random sequential
order~\cite{random1,random2,random3}. Firstly, an individual $i$ is
randomly selected who obtains the payoff $P_{i}$ according to the
above equations. Next, individual $i$ chooses one of its nearest
neighbors at random, and the chosen neighbor $j$ also acquires its
payoff $P_{j}$. Finally, individual $i$ adopts the neighbor $j$'s
strategy with the probability~\cite{update0}:

\begin{equation}
W(s_{i}\leftarrow s_{j})=\frac{1}{1+\exp[(P_i-P_j)/K]}, \label{7}
\end{equation}
where $K$ characterizes the noise introduced to permit irrational
choices. Following the previous studies~\cite{noise1,noise2}, we set
the noise level to be $K = 0.5$.

The key quantity for characterizing the cooperative behavior of the
system is the fraction of cooperators $\rho_{c}$ in the steady
state.  In our simulation, $\rho_{c}$ is obtained by averaging over
the last $10^{4}$ Monte Carlo steps (MCS) of the entire $10^{6}$
MCS. Each MCS consists of on average one strategy-updating event for
all individuals. Each data is obtained by averaging over 20
different realizations.

\section{Results}\label{sec: results}

\begin{figure}
\begin{center}
\scalebox{0.4}[0.4]{\includegraphics{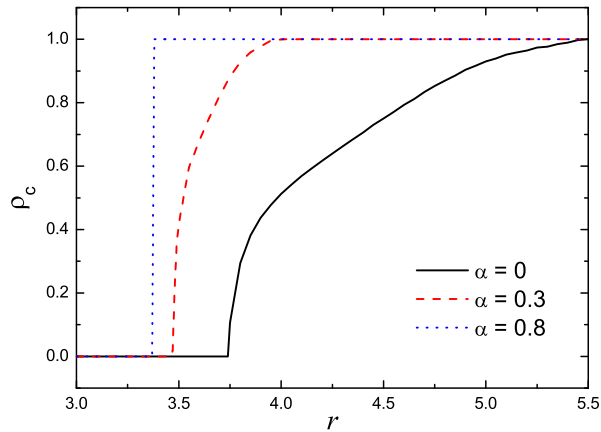}} \caption{(Color
online) The fraction of cooperators $\rho_{c}$ as a function of the
multiplication factor $r$ for different values of the punishment
fine $\alpha$. For small values of $\alpha$ (e.g., $\alpha=0$ or
$\alpha=0.3$), the dependence of $\rho_{c}$ on $r$ displays a
continuous phase transition. However, for large values of $\alpha$
(e.g., $\alpha=0.8$), the phase transition becomes discontinuous.}
\label{fig1}
\end{center}
\end{figure}

\begin{figure}
\begin{center}
\scalebox{0.4}[0.4]{\includegraphics{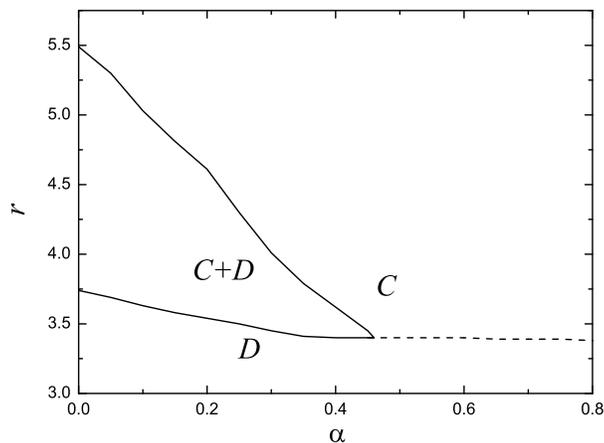}} \caption{Full
punishment fine-multiplication factor ($\alpha-r$) phase diagram.
There are three regions: full cooperators ($C$), full defectors
($D$), and the coexistence of cooperators and defectors ($C+D$).
Solid line denotes continuous phase transition, while dashed line
marks discontinuous phase transition. The region of $C+D$ disappears
when $\alpha>0.45$.} \label{fig2}
\end{center}
\end{figure}

Figure~\ref{fig1} shows the fraction of cooperators $\rho_{c}$ as a
function of the multiplication factor $r$ for different values of
the punishment fine $\alpha$. From Fig.~\ref{fig1}, one can see
that, for no punishment ($\alpha=0$) or a light punishment (e.g.,
$\alpha=0.3$), $\rho_{c}$ continuously increases from 0 to 1 as $r$
increases. However, for a severe punishment (e.g., $\alpha=0.8$),
there exists an abrupt transition point ($r\approx 3.38$ for
$\alpha=0.8$), at which $\rho_{c}$ suddenly jumps from 0 to 1. For a
fixed value of $r$, the payoff difference between the two
neighboring nodes is limited when the punishment is light, thus
individuals with the minority strategy can survive in the system if
their payoffs are larger than that of neighbors with the majority
strategy. However, for a severe punishment, the payoff difference
between the two different strategies is large, individuals with the
local majority strategy gain much more than those with the local
minority strategy. In this case, the coexistence of the two
different strategies becomes extremely difficult. Once a strategy
becomes dominated in the late stage of evolution, it will
determinately take over the whole system.

In Fig.~\ref{fig2}, we plot the full $\alpha-r$ phase diagram, which
can be divided into three regions: full cooperators ($C$), full
defectors ($D$), and the coexistence of cooperators and defectors
($C+D$). Different regions are separated by solid lines. One can see
that, the critical value of the multiplication factor $r$ below
which cooperators die out and the critical value of the
multiplication factor $r$ above which defectors disappear decrease
as the punishment fine $\alpha$ increases. At the same time, the
region of $C+D$ becomes narrower as the punishment fine $\alpha$
increases. Particulary, the region of $C+D$ disappears when
$\alpha>0.45$, indicating that cooperators and defectors cannot
coexist in the case of severe punishment.

\begin{figure}
\begin{center}
\scalebox{0.45}[0.45]{\includegraphics{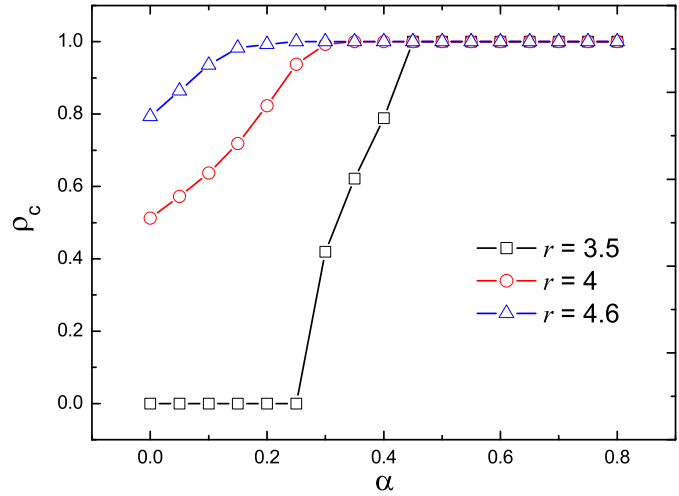}} \caption{(Color
online) The fraction of cooperators $\rho_{c}$ as a function of the
punishment fine $\alpha$ for different values of the multiplication
factor $r$. For each value of $r$, $\rho_{c}$ increases with
$\alpha$, indicating that severe punishment is in favor of
cooperation. } \label{fig3}
\end{center}
\end{figure}

Next, we study how the punishment fine affects the cooperation
level. From Fig.~\ref{fig3}, one can observe that for a fixed value
of the multiplication factor $r$, the fraction of cooperators
$\rho_{c}$ increases with the punishment fine $\alpha$, manifesting
that more severe punishment can better promote cooperation.

\begin{figure}
\begin{center}
\scalebox{0.4}[0.4]{\includegraphics{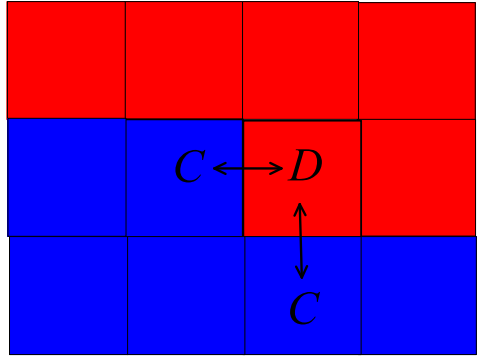}} \caption{(Color
online) Schematic presentation of the interface that separates
cooperators (blue) and defectors (red). The two leading elementary
processes that contribute the most to the change of the interface
are marked by arrows. The average payoff difference between the two
strategies is $P_{C}-P_{D}=3r/2+17\alpha/24-5$, which increases as
the punishment fine $\alpha$ increases.} \label{fig4}
\end{center}
\end{figure}

It has been known that the coevolution of cooperator clusters and
defector clusters plays an important role in spatial
games~\cite{cluster1,cluster2}. In Fig.~\ref{fig4}, we plot a
typical interface that separate clusters of the two competing
strategies. According to the analysis in Ref.~\cite{Chen}, the
leading invasions thereby are those which are marked with arrows.
The likelihood of other elementary processes is much smaller, and
hence their contribution to the evolution of clusters is negligible.
According to Fig.~\ref{fig4}, the average payoff difference between
the two strategies can be calculated as
\begin{equation}
P_{C}-P_{D}=\frac{3r}{2}+\frac{17\alpha}{24}-5.\label{e0}
\end{equation}
From Eq.~(\ref{e0}), one can find that the value of $P_{C}-P_{D}$
increases as $\alpha$ increases, implying that the cooperator
clusters can more effectively invade the defector clusters when
punishment is more severe.

\begin{figure*}
\begin{center}
\scalebox{0.8}[0.8]{\includegraphics{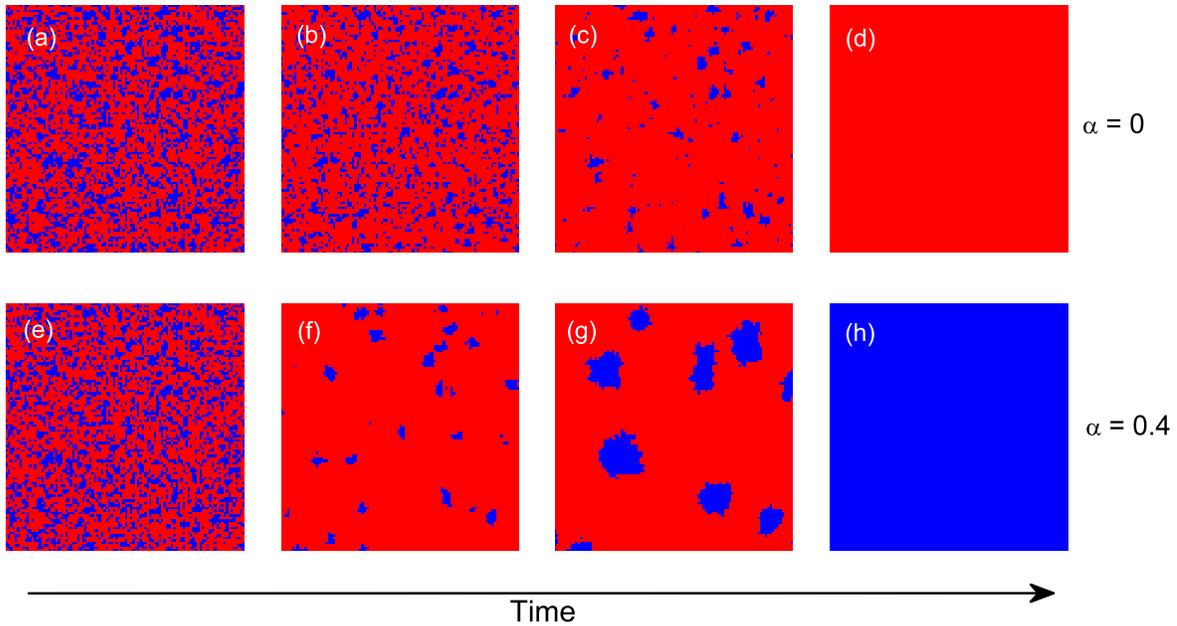}} \caption{(Color
online) Snapshots of typical distributions of cooperators (blue) and
defectors (red) at different time steps. To facilitate observe, we
use a smaller system size, that is, $200\times 200$ lattices. The
multiplication factor $r=3.7$. The punishment fine $\alpha$ for
(a)-(d) and (e)-(f) is 0 and 0.4 respectively. For $\alpha=0$,
cooperators gradually die out. For $\alpha=0.4$, initially the
number of cooperators greatly reduces and the survival cooperators
form some small clusters. With time the cooperator clusters continue
to expand and finally occupy the whole system.}\label{fig5}
\end{center}
\end{figure*}

To confirm the above analysis, we plot spatial strategy
distributions at different time steps. For $\alpha=0$, one can
observe that cooperators gradually die out, as shown in
Figs.~\ref{fig5}(a)-(d). For $\alpha=0.4$,  initially the density
cooperators decreases [see Fig.~\ref{fig5}(e) and
Fig.~\ref{fig5}(f)]. With time the cooperator clusters continue to
expand [see Fig.~\ref{fig5}(g)] and finally occupy the whole system
[see Fig.~\ref{fig5}(h)]. To more intuitively understand how the
punishment fine affects the evolution of clusters, we set initially
a giant cooperator (defector) cluster in the left (right) half of
square lattices. The multiplication factor is set to be $r=3.5$.
From Figs. 16(a)-(d), one can see that for the small value of
$\alpha$ (e.g., $\alpha=0.2$), the defector cluster gradually
invades the cooperator cluster and the original one big cooperator
cluster is divided into some small clusters. For the large value of
$\alpha$ (e.g., $\alpha=0.8$), the cooperator cluster continually
expands while the defector cluster gradually shrinks [see
Fig.~\ref{fig5}(e)-(h)]. Note that for $\alpha=0.2$, the interfaces
separating domains of cooperators and defectors become littery.
However, for $\alpha=0.8$, the boundary between the two competing
clusters remains smooth during the whole evolution. As pointed out
in Ref.~\cite{Perc} , noisy borders are beneficial for defectors,
while straight domain walls help cooperators to spread. From
Figs.~\ref{fig4}-\ref{fig5}, we can understand why cooperation can
be enhanced by enforcing more severe punishment.

\begin{figure}
\begin{center}
\scalebox{0.5}[0.5]{\includegraphics{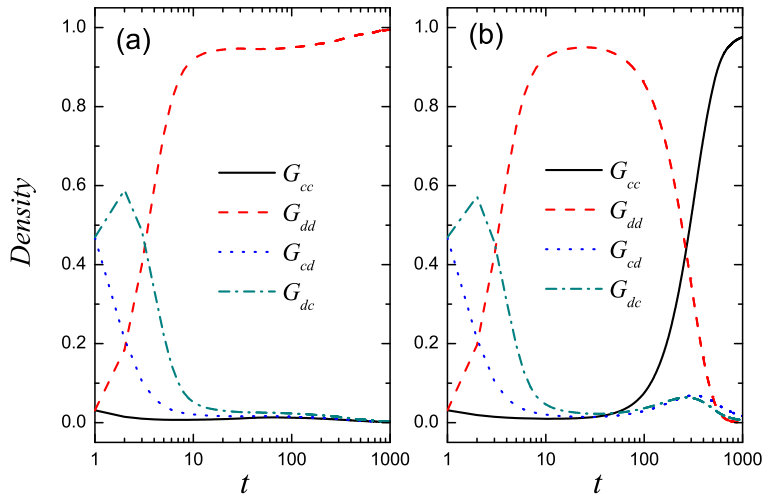}} \caption{(Color
online) Density of the four types of groups as a function of time
step $t$ when the multiplication factor $r=3.7$. The punishment fine
$\alpha$ for (a) and (b) is 0 and 0.4 respectively. $G_{cc}$
($G_{dd}$) denotes the PGG group which is occupied by full
cooperators (defectors), and $G_{cd}$ ($G_{dc}$) denotes the PGG
group in which cooperators (defectors) punish defectors
(cooperators). } \label{fig7}
\end{center}
\end{figure}

\begin{figure}
\begin{center}
\scalebox{0.4}[0.4]{\includegraphics{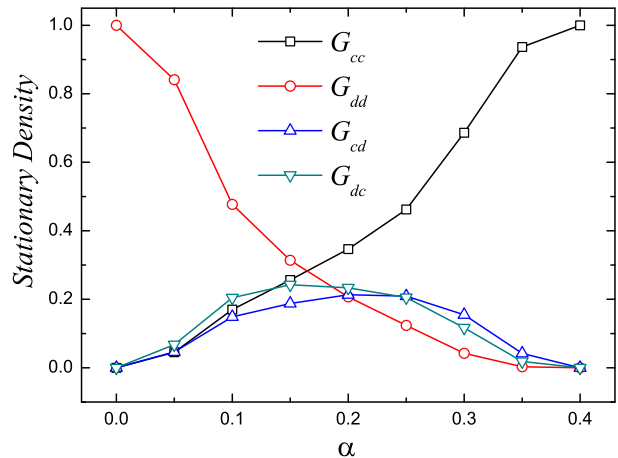}} \caption{(Color
online) Stationary density of the four types of groups as a function
of the punishment fine $\alpha$ when the multiplication factor
$r=3.7$. The stationary density of $G_{cc}$ ($G_{dd}$) groups
increases (decreases) as $\alpha$ increases. The stationary density
of $G_{cd}$ and $G_{dc}$ groups are maximized at moderate values of
$\alpha$. Note that for $\alpha<0.25$, the density of $G_{cd}$
groups is lower than that of $G_{dc}$ groups. Conversely, the
density of $G_{cd}$ groups is higher than that of $G_{dc}$ groups
when $\alpha>0.25$. } \label{fig8}
\end{center}
\end{figure}

We divide PGG groups into four types: $G_{cc}$ ($G_{dd}$) denotes
the group which is fully occupied by cooperators (defectors), and
$G_{cd}$ ($G_{dc}$) denotes the group in which cooperators
(defectors) punish defectors (cooperators). Figure~\ref{fig7} shows
the density of the four types of groups as a function of the time
step $t$ when the multiplication factor $r=3.7$. From
Fig.~\ref{fig7}(a), one can see that the density of $G_{cc}$
decreases to 0 as time evolves when the punishment fine $\alpha=0$.
However, for $\alpha=0.4$ [see Fig.~\ref{fig7}(b)], the density of
$G_{cc}$ firstly decreases and then increases to 1 as time evolves.
With time the density of $G_{dd}$ groups increases to 1 for
$\alpha=0$. However, for $\alpha=0.4$, the density of $G_{dd}$
groups firstly increases to be close to 1, and then decreases to 0.
For both of $\alpha=0$ and $\alpha=0.4$, the density of $G_{cd}$
gradually decreases to 0 as time $t$ increases. The density of
$G_{dc}$ firstly increases and then decreases to 0 as time evolves.
During the early evolution (i. e., $t<100$), the density of $G_{cd}$
groups is lower than that of $G_{dc}$ groups, indicating that
cooperators are more likely to be punished. We need to point out
that, although the larger value of $\alpha$ is more disadvantageous
for cooperators in the initial stage, it can help the surviving
cooperator clusters more effectively expand and finally take over
the whole system.

 Figure~\ref{fig8} shows the stationary density of
the four types of groups as a function of the punishment fine
$\alpha$ when the multiplication factor $r=3.7$. One can see that
the density of $G_{cc}$ groups increases with $\alpha$ while the
stationary density of $G_{dd}$ groups decreases as $\alpha$
increases. The stationary density of $G_{cd}$ or $G_{dc}$ groups
peaks at the moderate value of $\alpha$. It is interesting to note
that, for small values of $\alpha$ ($\alpha<0.25$), the stationary
density of $G_{cd}$ groups is lower than that of $G_{dc}$ groups. On
the contrary, for larger values of $\alpha$ ($\alpha>0.25$), the
stationary density of $G_{cd}$ groups is higher than that of
$G_{dc}$ groups, indicating that defectors become more likely to be
punished.

\section{Conclusions and discussions}\label{sec: conclusion}

To summarize, we have proposed a new mechanism of punishment in
which individuals with the majority strategy will punish those with
the minority strategy in a public goods game group. Note that in our
mechanism, an individual can play a dual role. He/She can be a
punisher in one group but meanwhile be punished in another group. We
find that the cooperation level increases with the punishment fine.
Compared to a light punishment, a severe punishment widens the
payoff gap between cooperators and defectors along the interface
that separates the two competing strategies, thus promotes the
expansion of cooperator clusters.

The interplay between opinion dynamics and evolutionary games has
received increasing attention. Szolnoki and Perc have found that,
suitably follow the majority strategy among neighbors can enhance
the cooperation level~\cite{inter1,inter2}. In our work, we propose
the punishment rule in which individuals with the majority strategy
are able to punish those with minority strategy. Thus, cooperators
and defectors could be punished by each other, depending on the
compositions in the groups. And then the pro-social and anti-social
punishments are both possible in our study. It is thus not intuitive
to predict whether cooperation can be promoted in the punishment
rule, especially when the complex structured interactions are
considered. Interestingly, we find that using the strategy of
punishing minority can increase the final cooperation level. In
particular, when the punishment fine is high, cooperators can
suddenly thrive and thus dominate the population from the outcome of
full defection. Together~\cite{inter1,inter2} and our work can
provide a deeper understanding of the impact of opinion dynamics on
the evolution of cooperation.

\begin{acknowledgments}
This work was supported by the National Natural Science Foundation
of China under Grants No. 61403083 and No. 61503062, and the
Fundamental Research Funds of the Central Universities of China.
\end{acknowledgments}


\begin{references}

\bibitem{rev1} Z. Wang, S. Kokubo, M. Jusup, J. Tanimoto, Phys. Life Rev. 14 (2015) 1.
\bibitem{rev2} Z. Wang, C. T. Bauch, S. Bhattacharyya, A. d'Onofrio, P. Manfredi, M. Perc, N. Perra, M. Salath\'{e}, D. Zhao, Phys. Rep. 664 (2016)
1.
\bibitem{rev3} M. Perc, J. J. Jordan, D. G. Rand, Z. Wang, S. Boccaletti, A.
Szolnoki, Phys. Rep. 687 (2017) 1.
\bibitem{1} M. Perc, J. G\'{o}mez-Garde\~{n}es, A. Szolnoki, L. M. Flor\'{\i}a, Y. Moreno, J R Soc. Interface 10 (2013) 20120997.
\bibitem{2} Z. Wang, S. Kokubo, J. Tanimoto, E. Fukuda, K. Shigaki, Phys. Rev. E 88 (2013) 042145.
\bibitem{2.1} Z. Wang, L. Wang, A. Szolnoki, M. Perc, Eur. Phys. J.
B 88 (2015) 124.

\bibitem{3} G. Szab\'{o}, C. Hauert, Phys. Rev. Lett. 89 (2002) 118101.
\bibitem{4} Z.-X. Wu, X.-J. Xu, Y. Chen, Y.-H. Wang, Phys. Rev. E
71 (2005) 037103.

\bibitem{5} A. Szolnoki, M. Perc, EPL 77 (2007) 30004.
\bibitem{7} A. Szolnoki, M. Perc, Phys. Rev. E 77 (2008) 011904.
\bibitem{8} Z.-X. Wu, Z. Rong, M. Z. Q. Chen, EPL 110 (2015) 30002.
\bibitem{9} M. Perc, A. Szolnoki, New J Phys. 10 (2008) 043036.
\bibitem{11} D. Helbing, W. Yu, Proc. Natl. Acad. Sci. USA 106 (2008) 3680.

\bibitem{13} X. Chen, A. Szolnoki, M. Perc. Phys. Rev. E 86 (2012) 036101.
\bibitem{14} M. Perc, EPL 75 (2006) 841.
\bibitem{15} A. Szolnoki, M. Perc, Phys. Rev. E 89 (2014) 022804.
\bibitem{16} D. Hao, Z. Rong, T. Zhou. Phys. Rev. E 91 (2015) 052803.
\bibitem{17} Z. Rong, Z.-X. Wu, D. Hao, M. Z. Q. Chen, T. Zhou, New J Phys. 17 (2015) 033032.
\bibitem{18} C. Wang, L. Wang, J. Wang, S. Sun, C. Xia, Appl. Math. Comput. 293 (2017) 18.
\bibitem{19} W.-X. Wang, J. Ren, G. Chen, B.-H. Wang, Phys. Rev. E 74 (2006) 056113.
\bibitem{20} W. Ye, W. Feng, C. L\"{u}, S. Fan, Appl. Math. Comput. 307 (2017) 31.
\bibitem{21} C.-Y. Xia, S. Meloni, M. Perc, Y. Moreno, EPL 109 (2015)
58002.
\bibitem{22} C. Shen, J.Lu, L. Shi, Appl. Math. Comput. 290 (2016)
201.
\bibitem{23} Z. Wang, M. Jusup, R.-W. Wang, L. Shi, Y.
Iwasa, Y. Moreno, J. Kurths, Sci. Adv. 3 (2017) e1601444.

\bibitem{the1} C. Hauert, S. De Monte, J. Hofbauer, K. Sigmund, Science 296 (2002) 1129.
\bibitem{the3} D. Helbing, A. Szolnoki, M. Perc, G. Szab\'{o}, New J
Phys. 12 (2010) 083005.
\bibitem{the4} D. Helbing, A. Szolnoki, M. Perc, G. Szab\'{o}, Phys. Rev.
E 81 (2010) 057104.
\bibitem{the5} A. Szolnoki, G. Szabo, L. Czako, Phys. Rev. E 84
(2011) 046106.
\bibitem{the6} A. Szolnoki, M. Perc, J Theor. Biol. 325 (2013) 34.
\bibitem{the7} A. Szolnoki, M. Perc, Phys. Rev. X 3 (2013) 041021.
\bibitem{the8} X. Chen, A. Szolnoki, M. Perc, Phys. Rev. E 92
(2015) 012819.
\bibitem{the9} H.-X. Yang, Z. Rong, Chaos Soliton Fractals 77 (2015) 230.


\bibitem{Szolnoki} A. Szolnoki, G. Szab\'{o}, M. Perc, Phys. Rev. E 83 (2011) 036101.

\bibitem{Perc} M. Perc, A. Szolnoki, New J Phys. 14 (2012) 043013.
\bibitem{Chen} X. Chen, A. Szolnoki, M. Perc, New J Phys. 16 (2014) 083016.
\bibitem{Wu} P. Cui, Z.-X. Wu, J Theor. Biol. 361 (2014) 111.



\bibitem{dissent1} N. Eisenberger, M. Lieberman, K. Williams, Science 290 (2003) 290.
\bibitem{dissent2} L. Somerville, T. Heatherton, W. Kelley, Nat.
Neurosci. 9 (2006) 1007.


\bibitem{random1} Z. Wang, A. Szolnoki, M. Perc, Sci. Rep. 3 (2013) 2470.
\bibitem{random2} Z. Wang, A. Szolnoki, M. Perc, J Theor. Biol. 349 (2014) 50.
\bibitem{random3} Z. Wang, L. Wang, M. Perc, Phys. Rev. E 89 (2014) 052813.
\bibitem{update0} G. Szab\'{o}, C. T\H{o}ke. Phys. Rev. E 58 (1998) 69.

\bibitem{noise1} A. Szolnoki, M. Perc, Phys. Rev. E 81 (2010) 057101.
\bibitem{noise2} A. Szolnoki, X. Chen, Phys. Rev. E 94 (2016) 042311.

\bibitem{cluster1} M. A. Nowak, R. M. May, Nature 359 (1992) 826.
\bibitem{cluster2} J. G\'{o}mez-Garde\~{n}es, M. Campillo, L. M. Flor\'{\i}a, Y. Moreno, Phys. Rev.
Lett. 98 (2007) 108103.

\bibitem{inter1} A. Szolnoki, M. Perc, J R Soc. Interface
12 (2015) 20141299.
\bibitem{inter2} A. Szolnoki, M. Perc, Sci. Rep. 6 (2016) 23633.

\end{references}
\end{document}